\documentclass[12pt,preprint]{aastex}

\usepackage[]{natbib}
\usepackage{graphicx}
\usepackage{amssymb}
\usepackage{color}
\usepackage{ulem}

\slugcomment{Ver. 2.1}

\input{macro.inp}

\shorttitle{}
\shortauthors{Hamaguchi et al.}

\begin{document}

\title{X-raying the Beating Heart of a Newborn Star: \\Rotational Modulation of High-energy Radiation from V1647~Ori}

\author{Kenji Hamaguchi\altaffilmark{1,2}, Nicolas Grosso\altaffilmark{3}, Joel H. Kastner\altaffilmark{4}, 
David A. Weintraub\altaffilmark{5}, Michael Richmond\altaffilmark{4}, Robert Petre\altaffilmark{6},
William K. Teets\altaffilmark{5}, David Principe\altaffilmark{4}}

\altaffiltext{1}{CRESST and X-ray Astrophysics Laboratory NASA/GSFC,
Greenbelt, MD 20771; Kenji.Hamaguchi@nasa.gov.}
\altaffiltext{2}{Department of Physics, University of Maryland, Baltimore County, 
1000 Hilltop Circle, Baltimore, MD 21250.}
\altaffiltext{3}{Observatoire Astronomique de Strasbourg, Universit\'e de Strasbourg,
CNRS, UMR 7550, 11 rue de l\textquoteright Universit\'e, 67000 Strasbourg, France}
\altaffiltext{4}{Laboratory for Multiwavelength Astrophysics, Rochester Institute of Technology, 54 Lomb Memorial Drive, Rochester, NY 14623.}
\altaffiltext{5}{Department of Physics \& Astronomy, Vanderbilt University, Nashville, TN 37235.}
\altaffiltext{6}{X-Ray Astrophysics Laboratory, NASA Goddard Space Flight Center, Greenbelt, MD 20771}

\begin{abstract}
We report a periodicity of $\sim$1 day in the highly elevated X-ray emission 
from the protostar V1647 Ori during its two recent multiple-year outbursts of mass accretion. This periodicity is indicative of protostellar rotation 
at near-breakup speed. Modeling of the phased X-ray light curve indicates 
the high-temperature ($\sim$50 MK), X-ray-emitting plasma, which is most 
likely heated by accretion-induced magnetic reconnection, resides in 
dense ($\gtrsim$5$\times$10$^{10}$~\UNITPPCC), pancake-shaped magnetic footprints 
where the accretion stream feeds the newborn star. The sustained X-ray periodicity of V1647 Ori 
demonstrates that such protostellar magnetospheric accretion configurations can be stable over timescales of years.
\end{abstract}

\section{Introduction}
Strong X-ray emission and collimated jets from newborn stars, so-called protostars, indicate that energetic magnetic activity plays 
an important role in star formation. However, the thick gaseous envelopes that obscure protostars at visual wavelengths complicate 
efforts to probe their innermost active regions.
Low-mass stars like the Sun in the formation phase (hereafter, young stellar objects or YSOs) gradually accumulate mass 
from the parent cloud before igniting nuclear fusion in the stellar core.
Matter falling from the cloud first forms a disk around the central star and
matter in the innermost disk gradually accretes onto it.
The young star's gravitational potential can accelerate infalling 
matter up to a few hundred \UNITVEL. When the accreting material collides 
with the stellar surface, it is shock-heated to a few MK; this 
thermalized plasma can emit soft ($E \lesssim1$ keV) X-rays \citep{Kastner2002b,Sacco2010}.

Matter cannot fall from the disk onto the central star unless it loses a significant fraction of its angular momentum.
Coupling of the large scale magnetic fields of the central star to 
those of the innermost disk is suspected to prompt the momentum transfer \citep[e.g., ][]{shu1996,hayashi1996}.
Magnetic reconnections triggered by the interactions may be involved in the ejection of a fraction of the infalling disk matter, 
along with most of the angular momentum, out of the system; 
the rest is accreted onto the central star.
Bipolar jets or collimated outflows seen ubiquitously outside of the YSO's envelopes
are likely launched by magnetohydrodynamic or magnetocentrifugal processes \citep[e.g.,][and references therein]{Reipurth2001}.

Although protostellar magnetospheric accretion models are widely 
accepted, and have gained support from observations of specific, 
relatively evolved objects \citep[e.g., classical T Tauri stars,][]{Donati2011a,Donati2011b}, the accretion geometry
-- and hence the validity and applicability of such 
models at earlier protostellar evolutionary stages -- 
remains uncertain in the case of younger, more deeply embedded objects
\citep{Johns-Krull2009}.
X-ray observations of 
rapidly accreting objects may offer a means to probe this magnetospheric 
accretion process.
The bulk of a star's mass is accreted during very early (observationally, 
Class 0 and I) phases, when the star is still deeply embedded in its 
parent cloud.
Such protostars tend to emit hard ($>$2 keV) X-rays with occasional rapid ($\lesssim$1~day) flares \citep{Imanishi2001}.
Some of the flares have been proposed as arising in large magnetic loops that connect the inner accretion disk 
and the stellar surface
(i.e., in a non-solar-type magnetic structure), which may be loaded by 
accreting material \citep{Tsuboi1998,Montmerle2000,Favata2005}.
However, systematic studies of X-ray-emitting pre-main sequence stars 
in the Orion nebula suggest that the occurrence of the largest magnetic 
loops, associated with major flare events, is not dependent on the 
presence of circumstellar disks
\citep{Getman2008c,Getman2008b,Aarnio2010}.

Over the last decade, a small number of YSOs have been found
to display sudden increases in mass accretion rate, by as much as a few orders of magnitude 
\citep{Hartmann1996}.
Such eruptive YSOs are crudely classified as either FU Ori (FUor) or EX Lupi (EXor) types, 
wherein the former generally display major outbursts over timescales of decades, and the latter generally display smaller, shorter-duration, outbursts. 
One such eruptive YSO, V1647 Ori --- which does not fit neatly into either the FUor or EXor eruptive class \citep[see discussion in][]{Teets2011} --- went into outburst during the period 2004--2006 and then, again, in 2008 (the latter eruption is evidently still ongoing).
This Class~I YSO \citep[][Principe et al. 2012, in prep.]{Muzerolle2005} offered the first direct
evidence that X-ray activity surges with increases in mass accretion activity \citep{Kastner2004,Kastner2006,Teets2011}.
During these eruptions, the X-ray luminosity of V1647 Ori increased by two orders of magnitude and the plasma temperature reached 
$\sim$50 MK \citep{Kastner2004,Grosso2005,Grosso2006,Kastner2006,Hamaguchi2010,Teets2011}.
Subsequently, other YSOs that experienced mass accretion outbursts were found to have similarly enhanced or strong X-ray activity 
\citep{Audard2010,Skinner2006, Skinner2009,Grosso2010}.
The gravitational potential of YSOs is not deep enough to thermalize plasma to such a high temperature;
yet the elevated levels of X-ray emission observed from YSOs undergoing accretion outbursts (such 
as V1647 Ori) strongly indicate an intimate connection between 
accretion activity and X-ray emission in these objects.
In order to determine where in the star/disk system the hard X-rays actually originate,
we must  ultimately understand the origin of the high energy activity 
during accretion outbursts. To this end, we have reanalyzed all X-ray 
observations of V1647 Ori in outburst, in search of temporal evidence 
that might reveal the site(s) and, perhaps, mechanism(s) responsible for 
its enhanced high-energy emission during these events.

\section{Data Sets}
Since the onset of the first outburst in 2004, V1647~Ori has been monitored 17 times
with three major X-ray observatories: 13 times with \CHANDRA\ \citep{Weisskopf2002}, 
three times with \XMM\ \citep{Jansen2001} and once with \SUZAKU\ \citep{Mitsuda2007}.
Six \CHANDRA\ observations during the first outburst and an \XMM\ observation during the second outburst
did not collect enough photons for timing analysis.
Table~\ref{tbl:obslogs} lists the 9 data sets used for the timing analysis and the 2 additional data sets added to the folded light curve plots
(Figs.~\ref{fig:periodic_lightcurve}, \ref{fig:periodic_lightcurve_flux}).
Hereafter, individual \CHANDRA, \XMM\ and \SUZAKU\ observations are designated CXO, XMM and SUZ, respectively,
subscripted with the year, month and day of the observation.

We reprocessed data with the following calibration versions: DS~7.6 or later for the \CHANDRA\ data, SAS~10.0.0 or later for the \XMM\ data and version~2.2.11.22 for 
the \SUZAKU\ data.
The basic analysis follows procedures described in earlier papers --- \CHANDRA~\citep{Kastner2006,Teets2011}, \XMM~\citep{Grosso2005}, and \SUZAKU~\citep{Hamaguchi2010}.
A notable departure from those procedures is that we converted event arrival times to the solar barycenteric time system,
which differs by up to 300~sec from the terrestrial time.
We generated background subtracted light curves for photons with energies in the range of 1$-$8 keV, binned into 2 ksec ($= \Delta T$) intervals,
using the HEAsoft\footnote{http://heasarc.gsfc.nasa.gov/docs/software/lheasoft/} analysis package.
We developed Python codes for the cross-correlation, chi-square and physical model studies.

\section{Period Search}
\label{sec:period}

\begin{figure}[t]
\plotone{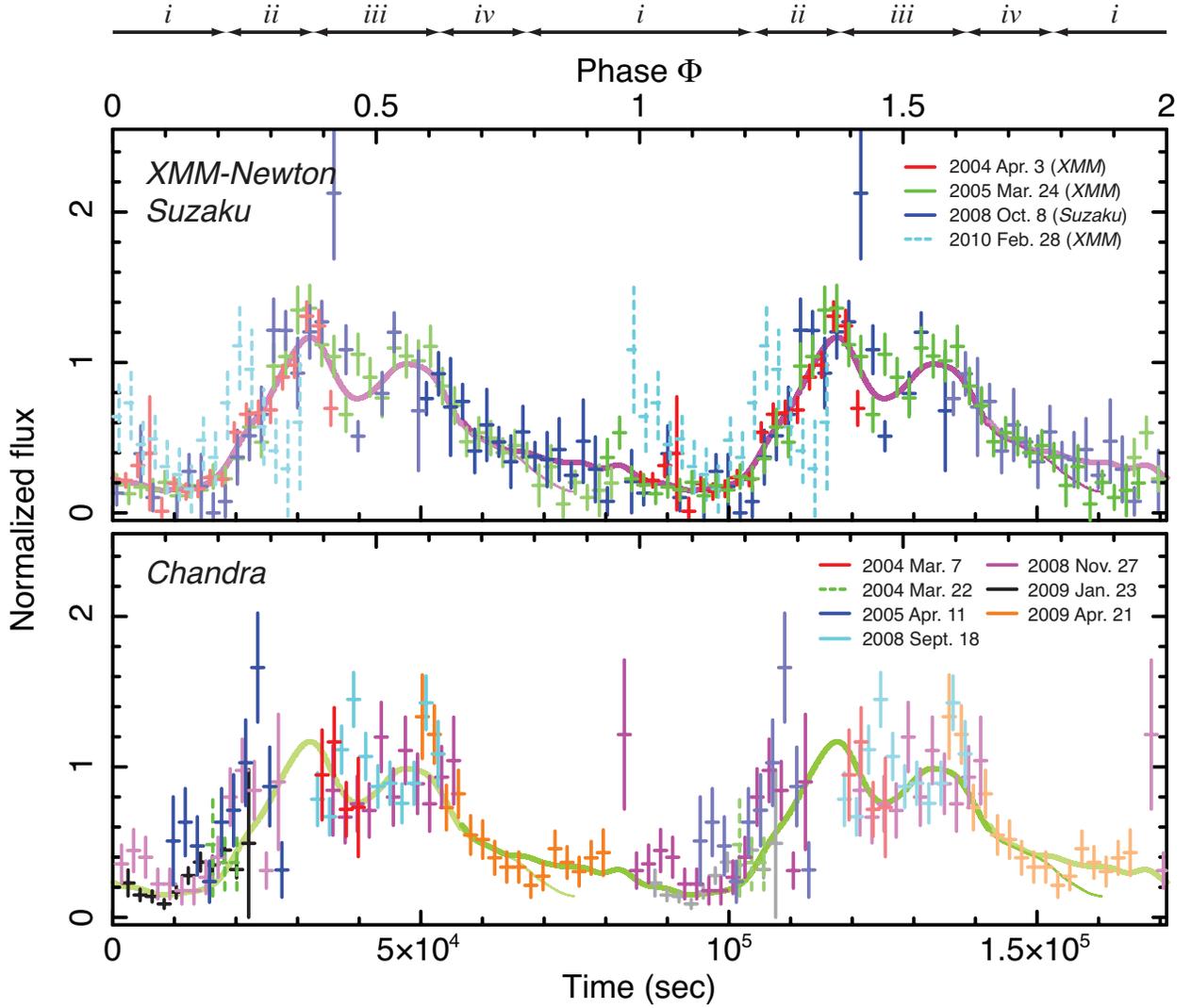}
\caption{
Light curves of V1647~Ori between 1$-$8~keV
obtained with the \XMM, \SUZAKU\ ({\it top}) and \CHANDRA\ ({\it bottom}) observatories.
These light curves are folded and normalized according to the best fit ephemeris
($f_{0} =$0.98929~day$^{-1}$, $\Phi_{gap} =-$0.383; see Appendix~\ref{sec:chisquare}).
The upper and lower horizontal axes in each panel show the rotational phase and time in second from the phase origin, respectively.
The phase origin is set at the middle of the flux fall and rise.
Light curves in thin colors are repeats of thick ones in the same colors.
Points with dotted error bars are not used for the best ephemeris search.
Solid purple and green lines in the {\it top} and {\it bottom} panels are the template light curves for the $\chi^{2}$ analysis.
Curves with narrow widths indicate phase intervals that have been disregarded, as per the algorithm described in Appendix~\ref{sec:chisquare}.
The arrows at the top depicts the phases of variation mentioned in Section \ref{sec:period}.
\label{fig:periodic_lightcurve}
}
\end{figure}

\begin{figure}[t]
\plotone{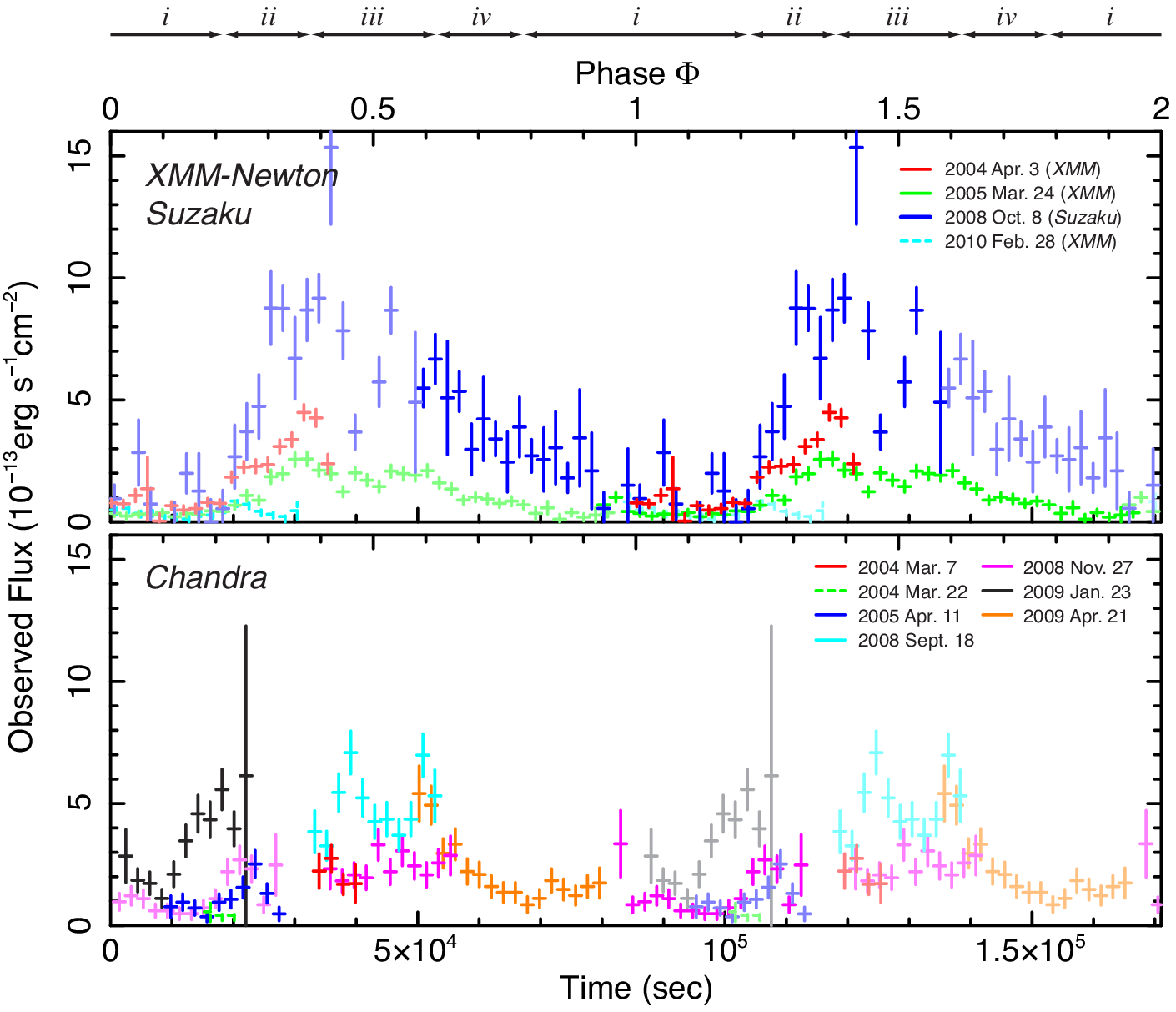}
\caption{
Same as Fig. \ref{fig:periodic_lightcurve}, but with axes in energy flux units.
\label{fig:periodic_lightcurve_flux}
}
\end{figure}

We identified strong similarities in eleven separate X-ray light curve observations of 
V1647 Ori obtained during the two outbursts with \CHANDRA, \XMM\ and \SUZAKU.
These flux variations --- an order of magnitude on timescales of hours --- superficially resemble stellar magnetic flares,
but there is reason to doubt this interpretation, given the spectral variation and frequency of flux rises \citep{Grosso2005,Kastner2006,Hamaguchi2010}.
Fig.~\ref{fig:periodic_lightcurve} displays all 11 light curves with time offset and flux normalization 
according to the $\chi^2$ study described below.
We first focus on the light curve with the longest duration, obtained with \XMM\ in 2005 (Fig.~\ref{fig:periodic_lightcurve}, {\it top}; 
the numbering scheme in the following sentence 
corresponds with the labels at the top of the figure).
The X-ray flux {\it i}) stays constant for $\sim$20~ksec, 
{\it ii}) rises by a factor of $\sim$5 in $\sim$15~ksec,
{\it iii}) keeps an elevated level for 30~ksec with marginal spikes and dips, 
and {\it iv}) falls gradually to the original flux level on a similar time scale.
The \XMM\ light curve in 2004 apparently matches with {\it i})$-${\it iii}), and the \SUZAKU\ light curve in 2008 starts from {\it iv}) and connects to {\it i})$-${\it iii}).
Although the \CHANDRA\ light curves (Fig.~\ref{fig:periodic_lightcurve}  {\it bottom}) are of more limited durations and have lower photon statistics, each of those light curves
also matches with one or more parts of profile {\it i})$-${\it iv}).

A cross-correlation analysis provides quantitative support to these qualitative similarities (see Appendix~\ref{sec:crosscor} for details).
The \XMM\ light curves in 2004 and 2005 correlate strongly (0.92)
when they align at their observation starts and the former is shifted backward by 6~ksec.
The \XMM\ light curve in 2005 and the \SUZAKU\ light curve in 2008 correlate strongly (0.82)
when they are folded by a period of 86~ksec and the \SUZAKU\ light curve shifts backward by 30~ksec.

Given the similarities of these light curves, we search for a more accurate period and set of phase shifts that match both the shapes and the timings of all the available light curves
(see Appendix~\ref{sec:chisquare} for details).
First, we generate a template light curve with a 1.23~day span by combining the \XMM\ and \SUZAKU\ light curves
that are shifted in time and normalized in flux based on the cross-correlation study.
We repeat the template light curve with a frequency below 1.45~day$^{-1}$, giving a phase offset between 0.0$-$1.0
and a flux normalization for each light curve to account for the long-term variation \citep{Teets2011}.
Based on our preliminary analysis, we also introduce a phase gap between the first and second outbursts.
We fit the template to all light curves with the least $\chi^{2}$ method and 
derive the minimum $\chi^{2}$ value of 317.9 [$\chi^{2}/d.o.f$ = 1.88 ($d.o.f.$ = 169)] at $f_{0} =$ 0.98929~day$^{-1}$ ($P =$87.3~ksec) and $\Phi_{gap} = -$0.383.
The best-fit period is close to the rotation period of the Earth; however, because none of the (space-based)
observatories whose data are analyzed here obtain data on a daily cadence,
this period cannot be an artifact of our observing protocol.  We therefore conclude that 
V1647 Ori displays periodic variation of its X-ray emission with a period of $\sim$1 day.

One obvious potential origin for the X-rays would be the rotation into and out of our line of sight of a localized region of X-ray plasma on the central star.
Rises and falls in the light curves would correspond to appearances and disappearances of the localized X-ray bright spot.
The flux transitions take $\sim$20\% of one cycle (Fig.~\ref{fig:lc_onecycle}, Appendix~\ref{sec:physmodel}),
suggesting that the spot has a significant size or height compared with the radius of the central star.

We assume a uniform circular spot with a tip-cut cone shape and simulate
an X-ray light curve for each combination of the spot radius, height, latitude and stellar inclination angle.
We find that no single spot can produce both the low flux interval ($\phi \sim$0.0$-$0.2, 0.8$-$1.0) and the high flux interval ($\phi \sim$0.4$-$0.6), so
we adopt a bipolar geometry and add a fainter spot at the opposite longitude and latitude of the brighter spot
whose shape is identical to that of the bright spot (see Appendix~\ref{sec:physmodel} for details).
The best-fit result ($\chi^2 =$98.2, $d.o.f. =$36) is obtained 
with a bright spot to faint spot brightness ratio of 5, 
with the spots having radii of 0.32~$R_{\ast}$ and heights of 0.01~$R_{\ast}$, 
and with the bright spot found at a stellar latitude of $\sim-$49\DEGREE.
The stellar inclination --- the tilt of the polar axis toward our line of sight --- is $\sim$68\DEGREE.
This model, shown in Fig.~\ref{fig:lc_onecycle}, reproduces the low and high flux levels and the rise and fall of the represented light curve well.
The possible dip at $\Phi=0.5$ may be reproduced by a partial eclipse of the bright spot by its own accretion flow or 
the disappearance of the faint spot behind the central star if it has a smaller size and higher latitude than those assumed in the fit above.
Excesses at $\Phi=$ 0.4 and 0.8 after the flux rise and fall 
may represent asymmetries of the spot shapes or the presence of additional smaller spots.
The best-fit for the inclination angle of the rotation axis is similar to the angle ($\sim$61\DEGREE) estimated from an infrared light echo study \citep{Acosta-Pulido2007}.

The one-cycle light curve can also be reproduced given plasma 
configurations that differ somewhat from the geometry described in Fig. 
3. For example, the faint hot spot --- which is required in the preceding 
model so as to reproduce the low flux levels during phases intervals 
$\Phi =0.0-0.2, 0.8-1.0$ ---
may shift toward the latitude and longitude directions with respect to the bright hot spot, or can be replaced
by a constantly visible plasma in the form, e.g., of a halo around the star.
Furthermore, the bright hot spot could really be a complex of multiple, smaller hot spots, instead of a uniform single spot.
However, in any plasma 
configuration, the flux increase during the phase interval $\Phi =0.2-0.8$
constrains the bright hot spot (or the envelope of multiple hot spots) 
to have the approximate size, height and latitude described above.

\section{Discussion}

The confidence ranges of the stellar inclination and the latitude of the bright spot (Fig.~\ref{fig:toymodel} {\it left})
exclude solutions involving a bright spot in the hemisphere facing us, a bright spot at a high latitude,
and a pole-on view of the system.
This result is consistent with modeling of strong fluorescent iron lines observed in \SUZAKU\ and \CHANDRA\ spectra \citep{Hamaguchi2010,Teets2011},
which suggests that a significant fraction of hard X-ray-emitting plasma is occulted from view.
The spot radius can be as large as $\sim$0.5$R_{\ast}$, 
while its height should be lower than $\sim$0.1$R_{\ast}$ (Fig.~\ref{fig:toymodel} {\it right}).
The spot --- shaped like a thin, extensive plate --- is similar to shape to the mass accretion footpoints of neutron stars, white dwarfs, and the Earth's aurorae.

No periodic variation such as that seen in the X-ray regime has been identified in optical or infrared observations of V1647~Ori, 
most likely because the optical and infrared emission mostly comes from the disk.
Based on a pre-outburst bolometric luminosity and stellar effective temperature,
\citet{Aspin2008b} roughly estimated the mass and radius of V1647~Ori at
$\sim$0.8\UNITSOLARMASS\ and $\sim$5\UNITSOLARRADIUS, respectively.
The stellar radius is slightly larger than the distance at which an orbiting body around a 0.8\UNITSOLARMASS\ star would have an orbital period of about one day.\footnote{Hereafter, we assume a stellar radius of 4 solar radii --- the approximate maximum radius that a 0.8\UNITSOLARMASS\ star with rotation period $\sim$1~day can maintain without breaking up.}
Thus, the central star must be rotating at a speed close to break-up velocity (i.e., rotating at the Keplerian velocity at the stellar radius).

Fig.~\ref{fig:toymodel} {\it right} also plots the product of the plasma density squared ($n^{2}$), the plasma filling factor ($\eta$), and the cube of the stellar radius ($R_{\ast}$),
using the plasma emission measure determined during the \SUZAKU\ observation in 2008.
There is no solution at log~${\eta}n^{2} (R_{\ast}/4R_{\odot})^{3}$ $\lesssim$2$\times$21~cm$^{-6}$.
Since $\eta\leqq$1 and $R_{\ast} \lesssim 4R_{\odot}$, we estimate $n \gtrsim$5$\times$10$^{10}$~\UNITPPCC.
Being only a lower limit, this result for density --- which is similar to those of the densest active stellar coronae 
\citep[e.g.,][]{Ness2004} --- may place V1647 Ori in the density regime inferred for the footpoints of free-fall accretion 
(as measured for isolated T Tauri stars via line ratios of helium-like ions; e.g.,
\citealt{Kastner2002b} for TW Hya; see also \citealt{Porquet2010}).
The magnetic field $B$ should be stronger than $B$~$\gtrsim$100~Gauss at the footpoints, 
in order to confine such a dense hot plasma
(i.e., if the magnetic pressure is stronger than the plasma pressure; plasma $\beta = n kT / (B^{2}/8\pi) < 1$).

The phase gap that we infer between the first and second outbursts may be caused by 
drift of the magnetic poles on the stellar surface or a change in the stellar rotational frequency.
For the latter case, we can replace the phase gap in our model ephemeris with a frequency derivative. In doing so, we
find a similarly good solution 
at a similar frequency with a small derivative
(see Appendix \ref{sec:chisquare}).

The observed X-ray variation of V1647 Ori can be naturally explained by rotational modulation of X-ray bright spots.
Coronal active regions can produce such rotational X-ray modulation \citep{Flaccomio2005}.
However, earlier studies \citep{Kastner2004,Kastner2006,Teets2011} 
indicate that large increases in X-ray flux from V1647 Ori during 
the outbursts are very closely correlated with the (accretion-driven) 
optical/infrared flux; therefore, these X-ray eruptions are most likely 
accretion-driven as well. The geometrical model described in \S\ref{sec:period} (Fig. \ref{fig:lc_onecycle}) 
supports such a model, in that it indicates that the X-ray-emitting 
plasma lies at or very near the foot-points of mass accretion streams at 
the stellar surface.

Since the profile of the X-ray light curves did not change remarkably between the observations,
the intrinsic X-ray luminosity of the hot spots, that is, the accretion-induced magnetic reconnection activity, varies on a timescale of a week or longer.
Given the large overall variation in the amplitude of the X-ray flux from outburst to outburst (and even during outbursts) (see Fig.~\ref{fig:periodic_lightcurve_flux}), 
it is evident that the large-scale magnetic field configuration of the V1647 Ori star/disk system is preserved, even as the protostar 
undergoes dramatic changes in accretion rate. 
We suggest two possible mechanisms that may generate this condition:
({\it i}) the mass accretion flow stably disrupts the magnetic fields at the footpoint \cite[e.g.,][]{Brickhouse2010}, or
({\it ii}) the rotational shear between the star and the disk continuously twists the stellar bipolar magnetic fields
\citep[e.g.,][see also Fig.~\ref{fig:cartoon}]{Goodson1997}.

More evolved YSOs also have faint, hard X-ray emission from hot plasma.
This emission has usually been explained as due to a blend of emission from multiple micro-flares \citep[e.g., ][]{Caramazza2007}.
Our result demonstrates that the mass accretion activity also can generate hot plasma constantly by sustained magnetic reconnection.
The same mechanism may also operate on those YSOs with weaker mass accretion activity.

\section{Conclusion}

We have discovered rotational modulation of X-ray emission from the Class I protostar V1647 Ori via analysis of 11 X-ray light curves 
obtained with the \CHANDRA, \XMM\ and \SUZAKU\ observatories during this YSO's two recent mass accretion outbursts. 
Based on a cross-correlation study and period search, we determined a rotational period of $\sim$1 day, with either a phase gap apparent 
between the two outbursts or frequency variation. 
The single-cycle light curve can be successfully reproduced by emission from two hot spots on opposite poles on the stellar surface. 
The star rotates with a period of $\sim$1 day, close to the break-up velocity for a 0.8~\UNITSOLARMASS\ star with a radius of $\sim$4~\UNITSOLARRADIUS. 
The hot spots likely cover significant fractions of the stellar surface, while the height ($\sim$0.01~$R_\ast$) may be negligible compared with the stellar radius. 
The hot spot size and the plasma emission measure indicate relatively high plasma density ($\gtrsim$5$\times$10$^{10}$~\UNITPPCC), 
also pointing to an origin in accretion hot spots.
This result clearly demonstrates that hard X-ray-emitting plasma can be present in long-lived accretion footprints at the surfaces of protostars, 
and thereby constrains the geometry of magnetospheric accretion in early (Class I) protostellar evolutionary stages.

\acknowledgments

This work is performed while K.H. was supported by the NASA's Astrobiology Institute (RTOP 344-53-51) 
to the Goddard Center for Astrobiology (Michael J. Mumma, P. I.).
J.K.'s research on X-rays from erupting YSOs is supported by NASA/GSFC \XMM\ Guest Observer grant
NNX09AC11G to RIT.
This research has made use of data obtained from the High Energy Astrophysics Science Archive
Research Center (HEASARC), provided by NASA's Goddard Space Flight Center.

Facilities: \facility{\CHANDRA~(ACIS)}, \facility{\XMM~(EPIC)}, \facility{\SUZAKU~(XIS)}

\appendix

\section{Cross-Correlations of the Light Curves}
\label{sec:crosscor}

We cross-correlate light curves of XMM$_{\rm 040404}$, XMM$_{\rm 050324}$, and SUZ$_{\rm 081008}$, which have good photon statistics.
We define the cross-correlation index $r$,
\begin{equation}
r = \frac{\sum\limits_{i} [(x_i - \bar{x})(y_{i-d} - \bar{y})]}{\sqrt{\sum\limits_{i} (x_{i} - \bar{x})^{2}}\sqrt{\sum\limits_{i} (y_{i-d} - \bar{y})^{2}}}\\
\end{equation}
where $x_{i}$ and $y_{i}$ are count rates of the $i$-th bins of two light curves, 
$d$ is a delay in units of 2~ksec time bins,
$\bar{x}$ and $\bar{y}$ are averages of bins that contributes to the cross-correlation.
The index $r$ ranges between $-1$ and 1. Two light curves are identical when $r = 1$.

We first cross-correlate XMM$_{040404}$ ($x_{i}$) with XMM$_{050324}$ ($y_{i}$) (Fig.~\ref{fig:crosscorr} {\it left}).
There is a strong correlation of $r = 0.92$ at $d = 3$ (6~ksec).

Since SUZ$_{081008}$ and XMM$_{050324}$ have similar observing durations,
the number of bins that contributes to the cross-correlation becomes smaller at larger delays.
The $r$ index fluctuates strongly in these regions.
We thus require the number of contributing bins to be at least 10.
When SUZ$_{081008}$ ($x_{i}$) is shifted backward against XMM$_{050324}$ ($y_{i}$),
there is a strong correlation of $r = 0.91$ at $d = 28$ (56~ksec).
When XMM$_{050324}$ ($x_{i}$) is shifted backward against SUZ$_{081008}$ ($y_{i}$),
there is a strong correlation of $r = 0.81$ at $d = 15$ (30~ksec).
This result indicates that these two light curves folded at a certain period also match well.
We thus fold both light curves by 40$-$46 bins (80$-$92~ksec),
average overlapped bins with weighted mean values, and cross-correlate them.
We found that the $r$ index is a maximum of 0.82 when these light curves are folded by 43 bins (86~ksec) and 
XMM$_{050324}$ ($x_{i}$) is shifted backward by $d = 30$ (60~ksec) against SUZ$_{081008}$ ($y_{i}$) (Fig.~\ref{fig:crosscorr} {\it right}).

\section{Search for the Best Ephemeris with the $\chi^{2}$ Test}
\label{sec:chisquare}

The cross-correlation tests the similarity of two light curves,
but it does not consider the time interval between them.
We therefore search for an ephemeris that satisfies both shapes and timings of all the X-ray light curves, including those detected with \CHANDRA.

We first generate a template light curve from the XMM$_{040404}$, XMM$_{050324}$ and SUZ$_{081008}$ light curves.
Based on the cross-correlation study in Appendix~\ref{sec:crosscor},
we shift XMM$_{040404}$ and XMM$_{050324}$ backward by 18 and 15 bins (36 and 30 ksec) against SUZ$_{081008}$, respectively.
We normalize these light curves based on the average count rates of overlapped bins
and average them at weighted mean values.
After artificially increasing the time resolution of the averaged light curve by 100 times with linear interpolation,
we smooth it with a Gaussian function with $\sigma$ = 2~ksec so as to minimize statistical noise.
The resulting template light curve [$L_{temp}(t_{i})$] spans a time period of 1.23~day.

We define the ephemeris as:
\begin{eqnarray}
\Phi(T) &=& \Phi_{0} + f_{0}~(T-T_{0}) + \Phi_{gap}H(T-T_{gap})
\label{ephemeris_phase_gap}
\end{eqnarray}
where $f_{0}$ is frequency, $T_{0}$ is the time origin, $\Phi_{0}$ is the phase at $T = T_{0}$, 
$\Phi_{gap}$ is the phase gap, $H$ is the unit step function and 
$T_{gap}$ is the time of the phase gap.
We then assign phases, $\Psi(t_{i})$, to the template light curve according to
\begin{eqnarray}
\Psi(t_{i}) &= & f_{0}~t_{i}
\end{eqnarray}
and discard bins at $\Psi(t_{i}) \geq 1$.

We sum bins of the template light curve within the corresponding phase range of each bin of a measured light curve, that is,
\begin{eqnarray}
I_{temp}(T_{i}) &= & \sum_{j=m}^{n} L_{temp}(t_{j})
\end{eqnarray}
where $m$ and $n$ satisfy
\begin{eqnarray}
\Phi'(T_{i}) &=& \Phi(T_{i}) -  \lfloor\Phi(T_{i})\rfloor\\
\Psi(t_{m}) &\leqq& \Phi'(T_{i}) - \frac{\Delta \Phi}{2} < \Psi(t_{m+1})\\
\Psi(t_{n}) &\leqq& \Phi'(T_{i}) + \frac{\Delta \Phi}{2} < \Psi(t_{n+1})
\end{eqnarray}
where $\lfloor x \rfloor$ is the floor function and $\Delta \Phi = f_{0}\Delta T$.
We then normalize $I_{temp}(T_{i})$ for each measured light curve by the normalization factor $n_{obs}$ that gives a minimum $\chi^{2}$ value,
and derive a $\chi^{2}$ value for each set of $\Phi_{0}, \Phi_{gap}$, and $f_{0}$.
\begin{eqnarray}
n_{obs} &=& \frac{\sum L_{obs}(T_{i})I_{temp}(T_{i})/\Delta L_{obs}(T_{i})^{2}}{\sum I_{temp}(T_{i})^{2}/\Delta L_{obs}(T_{i})^{2}}\\
\chi^{2} &=& \sum_{obs} \sum_{i} (\frac{L_{obs}(T_{i}) - n_{obs} I_{temp}(T_{i})}{\Delta L_{obs}(T_{i})})^{2}
\end{eqnarray}
We search for the minimum $\chi^{2}$ value in the ranges of $0 \leqq \Phi_{0}, \Phi_{gap} < 1$ and $f_{0} \geqq$ 0.82~day$^{-1}$.
We find a minimum $\chi^{2}$ of 317.93 [$\chi^{2}/d.o.f.$ = 1.88~($d.o.f. =$169)] at $f_{0} =$ 0.98929~day$^{-1}$ ($P =$87.3~ksec)
and $\Phi_{gap} = -$0.383.
The best ephemeris is hence expressed as
\begin{eqnarray}
\Phi(T) &= & 0.98929~(T-13453.09075) -0.383~H(T-14300)
\label{ephemeris_gap}
\end{eqnarray}
where $T$ is the truncated Julian date.
In this formula, we re-define the $\Phi$ origin at an intermediate point between the fall and rise timings of the single cycle light curve (see Appendix~\ref{sec:physmodel}).

The phase gap is empirical. The phase change may instead be explained by a frequency variation. 
We hence modify equation (\ref{ephemeris_phase_gap}) to allow for a frequency derivative.
\begin{eqnarray}
\Phi(T) &=& \Phi_{0} + f_{0}~(T-T_{0}) + f_{1}~(T-T_{0})^{2} / 2
\label{ephemeris_freq_deriv}
\end{eqnarray}
We find a similarly good fit of $\chi^{2} = 317.22$ at
$f_{0} =$0.98932~day$^{-1}$ and $f_{1} = -$1.89$\times10^{-6}$~day$^{-2}$ at $T_{0} = 14747.6758097$~day.
Figs. \ref{fig:periodic_lightcurve_pdot} and \ref{fig:periodic_lightcurve_pdot_flux} show light curves folded with this ephemeris.

We generate a single cycle light curve from XMM$_{040404}$, XMM$_{050324}$ and SUZ$_{081008}$.
We fold these light curves with equation (\ref{ephemeris_gap}) and
bin all light curves with $\Delta T =$1984.9 second, so that one whole light curve requires exactly 44 bins.
We normalize and average these light curves in a manner identical to that which produced the template light curve.

\section{Physical Model}
\label{sec:physmodel}

We consider an X-ray point source sitting at longitude~$\theta$, latitude~$\phi$ and height~$h$ from the stellar surface,
viewed from the rotation axis at inclination angle $i$ (see Fig.~\ref{fig:geometry_point} {\it left}).
We define the $\theta$ origin as the longitude that crosses the opposite side of the central star from the Earth at $\Phi = 0$.
The point source appears in view at a longitudinal difference $\Delta \theta_{e}$ from this ($\theta$=0) reference point (Figs.~\ref{fig:geometry_point} {\it right})
that satisfies:
\begin{eqnarray}
cos \Delta\theta_{e} &=& \frac{cos i~sin\phi + \sqrt{1-(\frac{R_{\ast}}{R_{\ast}+h})^2}}{sin i~cos \phi}
\end{eqnarray}
The normalized X-ray light curve of the point source is:
\begin{eqnarray}
F_{\rm point}(\Phi) &= & 0~~(0 \leqq \Xi < \Phi_{e})\\
		        &=  & 1~~(\Phi_{e} \leqq \Xi \leqq 1 - \Phi_{e})\\
		        &=  & 0~~(1-\Phi_{e} < \Xi < 1)
\end{eqnarray}
where $\Xi = \Phi+\frac{\theta}{2\pi} - \lfloor\Phi+\frac{\theta}{2\pi}\rfloor$ and $\Phi_{e} = 2\pi\Delta\theta_{e}$.

X-ray spectra of V1647~Ori indicate that the X-ray plasma is optically thin \citep{Kastner2006,Grosso2005,Hamaguchi2010,Kastner2004,Teets2011}.
This means that any portion of the X-ray plasma with a finite size can be seen once the portion emerges from the stellar rim, such that
the observed flux at phase $\Phi$ is the integrated emission from the plasma emerging above the stellar rim, i.e.,
\begin{eqnarray}
F(\Phi, i) &= & \int \int \int F_{\rm point}(\Phi, \theta, \phi, h, i) S(\theta, \phi, h) (R_{\ast}+h)^{2} cos\phi d\theta d\phi dh
\label{eqn:genral_plasma_integral}
\end{eqnarray}
where $S(\theta, \phi, h)$ is the X-ray source distribution of the plasma.

For simplicity, we assume a uniform plasma with a conical shape with angular radius $r_{\ast}$ and height $h'$,
standing upside down on the stellar surface, such that the base position of the cone is "exposed" (Fig.~\ref{fig:geometry_cone}).
It is observed at unit flux when in view, such that the source distribution is described as
\begin{eqnarray}
S(\theta, \phi, h) & = & 1/V~~(inside)\\
			 & = & 0~~(outside)
\end{eqnarray}
where $V$ is the plasma volume.
We define the longitude and the latitude of the cone axis as $\theta'$ and $\phi'$, respectively.
A narrow latitudinal strip at $\phi' + \Delta\phi'$ $(|\Delta\phi'| \leq r_{\ast})$ has a half width $\Delta\theta_{s}$ that satisfies
\begin{eqnarray}
cos\Delta\theta_{s} &= & \frac{cos~r_{\ast} - sin(\phi'+\Delta\phi')sin\phi'}{cos(\phi'+\Delta\phi')cos \phi'}
\end{eqnarray}
We numerically integrate equation~(\ref{eqn:genral_plasma_integral}) for this plasma, i.e.,
\begin{eqnarray}
F_{\rm cone}(\Phi, \theta', \phi', r_{\ast}, h', i) &= & \int_{0}^{h'} \int_{\phi'-r_{\ast}}^{\phi'+r_{\ast}} \int_{\theta'-\Delta\theta_{s}}^{\theta'+\Delta\theta_{s}} \frac{F_{\rm point}(\Phi, \theta, \phi, h, i) (R_{\ast}+h)^{2} cos\phi}{V} d\theta d\phi dh
\end{eqnarray}

\begin{figure}[t]
\plotone{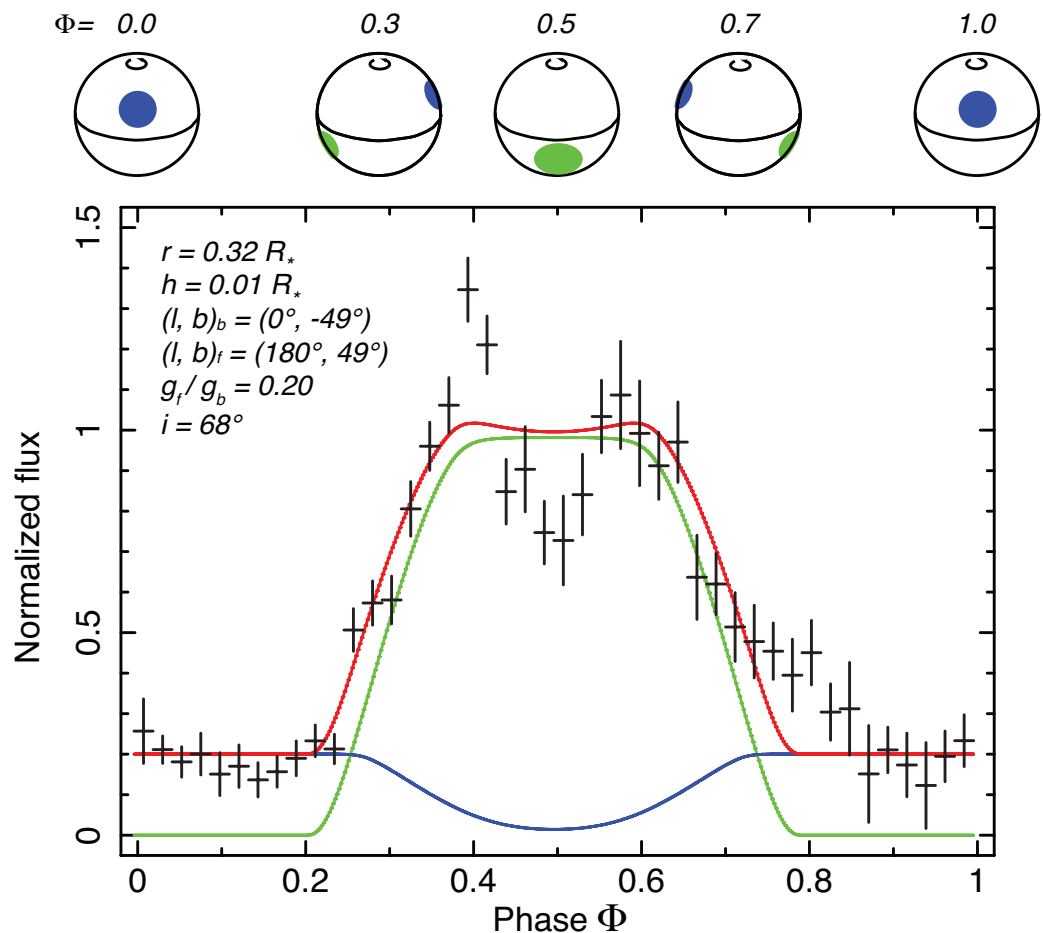}
\caption{
Single cycle light curve created from the \XMM\ and \SUZAKU\ light curves
overlaid with the best-fit result of the symmetric bipolar spot model ({\it red}: total, {\it green}: bright spot, {\it blue}: faint spot).
The pictures on the top depict locations of hot spots at corresponding phases.
See Appendix~\ref{sec:physmodel} for the numbers and letters in the label.
The subscripts b and f stand for the bright and faint spots, respectively.
\label{fig:lc_onecycle}
}
\end{figure}

The low and high flux phases in the single cycle light curve cannot be reproduced by any single spot.
We therefore assume two spots with identical shapes in a dipole geometry, sitting on opposite sides of the central star.
The X-ray light curve is, then, expressed as 
\begin{eqnarray}
F & = & g_{b} F_{\rm cone}(\Phi, \theta', \phi', r_{\ast}, h', i) + g_{f} F_{\rm cone}(\Phi, \theta'+\pi, -\phi', r_{\ast}, h', i)
\end{eqnarray}
where $g_{b}$ and $g_{f}$ are un-occulted fluxes of the bright and faint spots, respectively.
The $\chi^{2}$ value is a minimum (98.2) when $\phi' = -$49\DEGREE,
$r_{\ast} = 18$\DEGREE, $h' =$0.01$R_{\ast}$, $i = $68\DEGREE\ and $g_{f}/g_{b}=0.20$.
In this fit, we fixed the longitude $\theta'$ at 0, considering the definition of the $\Phi$ origin.
Fig.~\ref{fig:lc_onecycle} illustrates this best-fit model.

\begin{figure}[t]
\plotone{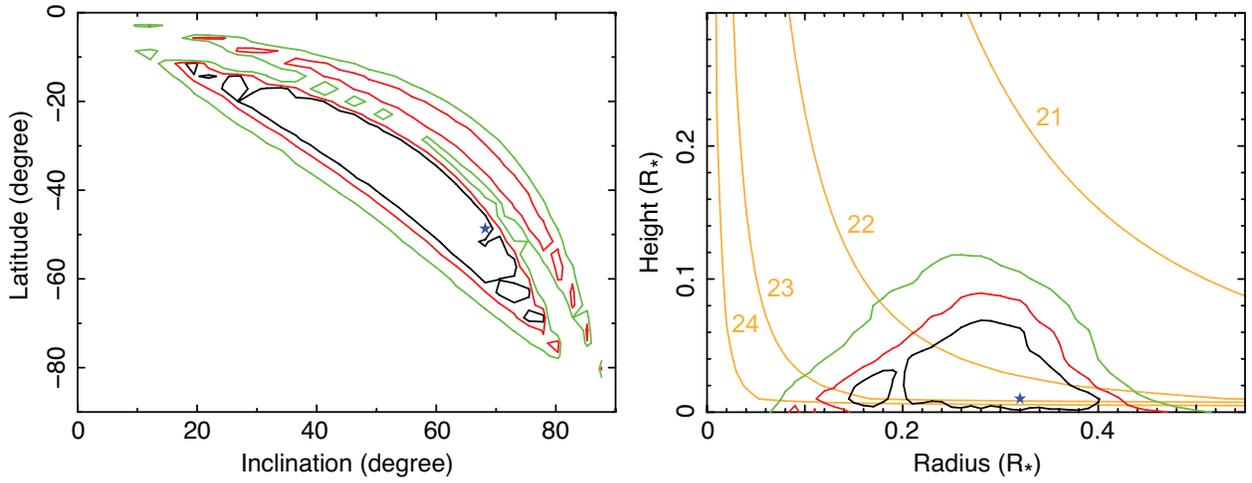}
\caption{Confidence ranges of parameter values obtained from model fitting (Fig.~\ref{fig:lc_onecycle}) ---
{\it left}: the stellar viewing inclination vs. the latitude of the bright spot;
{\it right}: the radius vs. height of the hot spots, in stellar radii.
In both panels, the black, red and green contours show confidence levels of 66\% ($\Delta\chi^{2} =$2.3),
90\% ($\Delta\chi^{2} =$4.6) and 99\% ($\Delta\chi^{2} =$9.2), respectively.
The blue stars show the best-fit result.
The right panel also shows contours of log~$\eta n^{2} (R_{\ast}/4R_{\odot})^{3} =$ 21, 22, 23, and 24 in 
solid orange lines as obtained from modeling \SUZAKU\ data \citep[][see Appendix~\ref{sec:physmodel} for the derivation.]{Hamaguchi2010}
\label{fig:toymodel}
}
\end{figure}

\begin{figure}[t]
\plotone{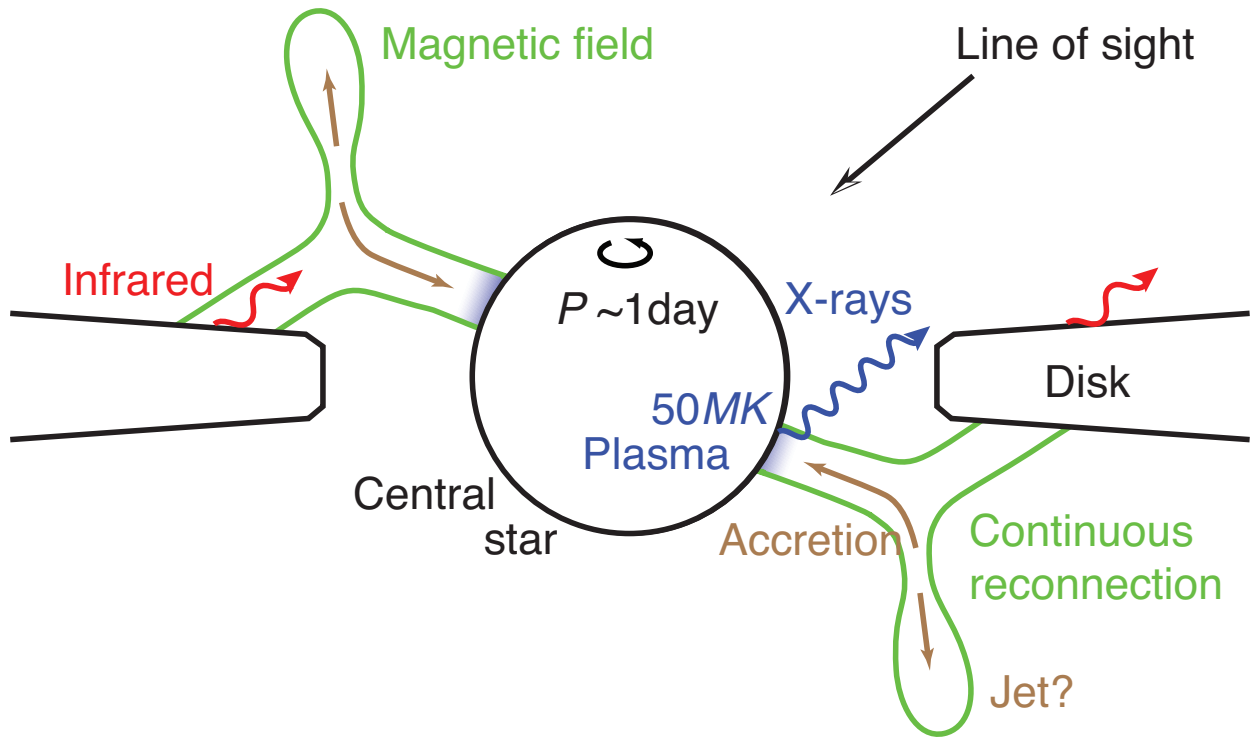}
\caption{A possible geometry of the V1647 Ori system (mechanism {\it ii}).
Differential rotation between the star and the disk shears the stellar bipolar magnetic fields, and
the magnetic fields twist and continuously reconnect.
Matter, accelerated to $v \thickapprox$2000 \UNITVEL\ by the magnetic reconnection,
collides with the stellar surface, thermalizes to \KT~$\sim$4~keV ($T$~$\sim$50 $MK$) and emits hard X-rays \citep[See e.g., the equation (9) of][]{Calvet1998}.
The opposite magnetic pole may have lower mass accretion and emit weaker X-rays.
\label{fig:cartoon}
}
\end{figure}

\begin{figure}[t]
\plotone{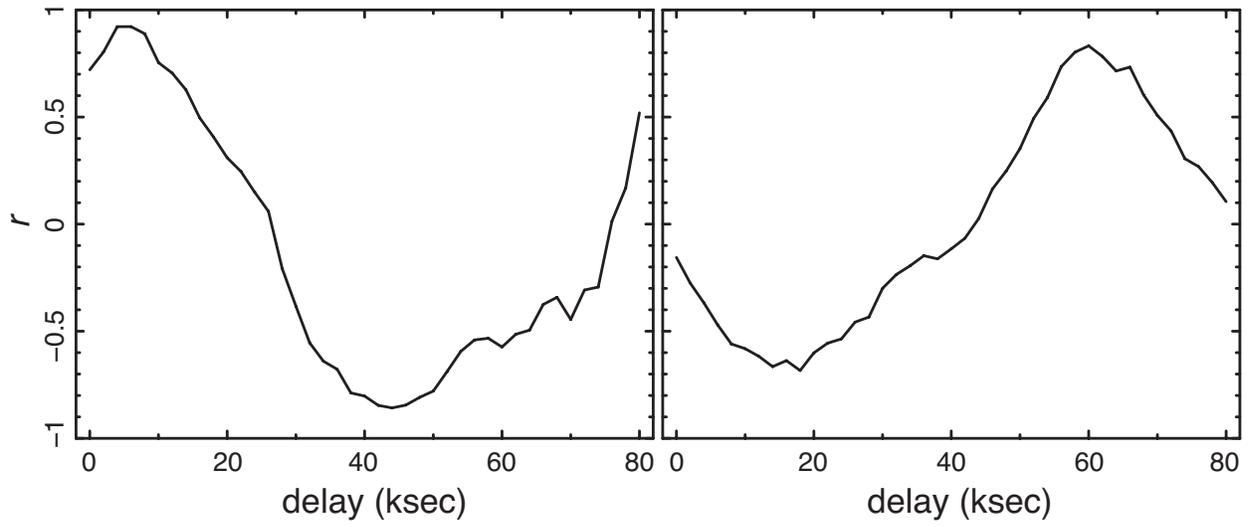}
\caption{Cross-correlation of XMM$_{050324}$ with XMM$_{040404}$ ({\it left})
and SUZ$_{081008}$ ({\it right}) when these light curves are folded by 43~bins (86~ksec).
\label{fig:crosscorr}
}
\end{figure}

\begin{figure}[t]
\plotone{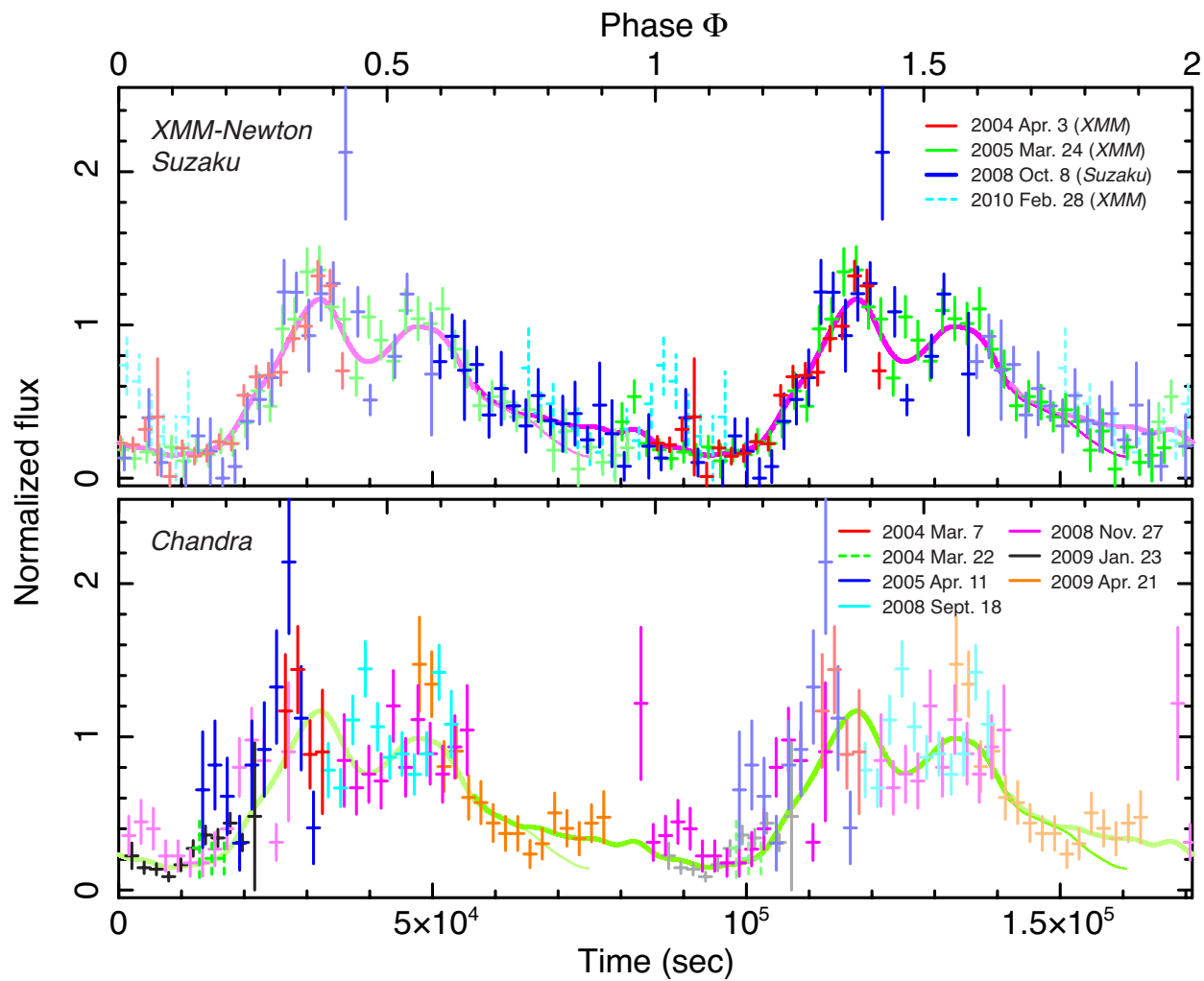}
\caption{Normalized light curves ({\it top}: \XMM, \SUZAKU, {\it bottom}: \CHANDRA) between 1$-$8~keV 
folded with the ephemeris in the equation (\ref{ephemeris_freq_deriv}), which assumes a frequency derivative instead of a phase gap.
See the caption of Fig.~\ref{fig:periodic_lightcurve} for the other details.
\label{fig:periodic_lightcurve_pdot}
}
\end{figure}

\begin{figure}[t]
\plotone{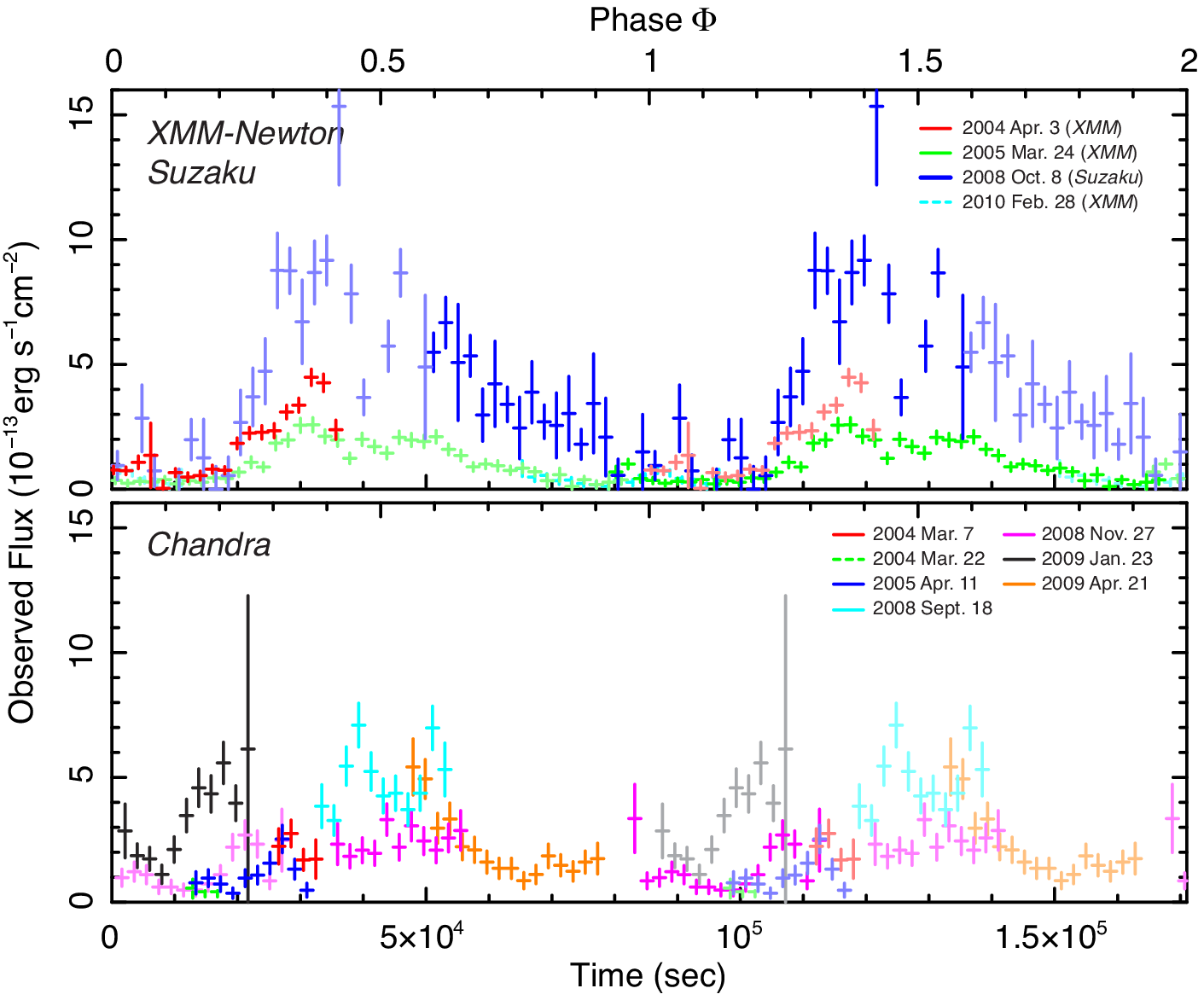}
\caption{Same as Fig. \ref{fig:periodic_lightcurve_pdot}, but with axes in energy flux units.
\label{fig:periodic_lightcurve_pdot_flux}
}
\end{figure}

\begin{figure}[t]
\plotone{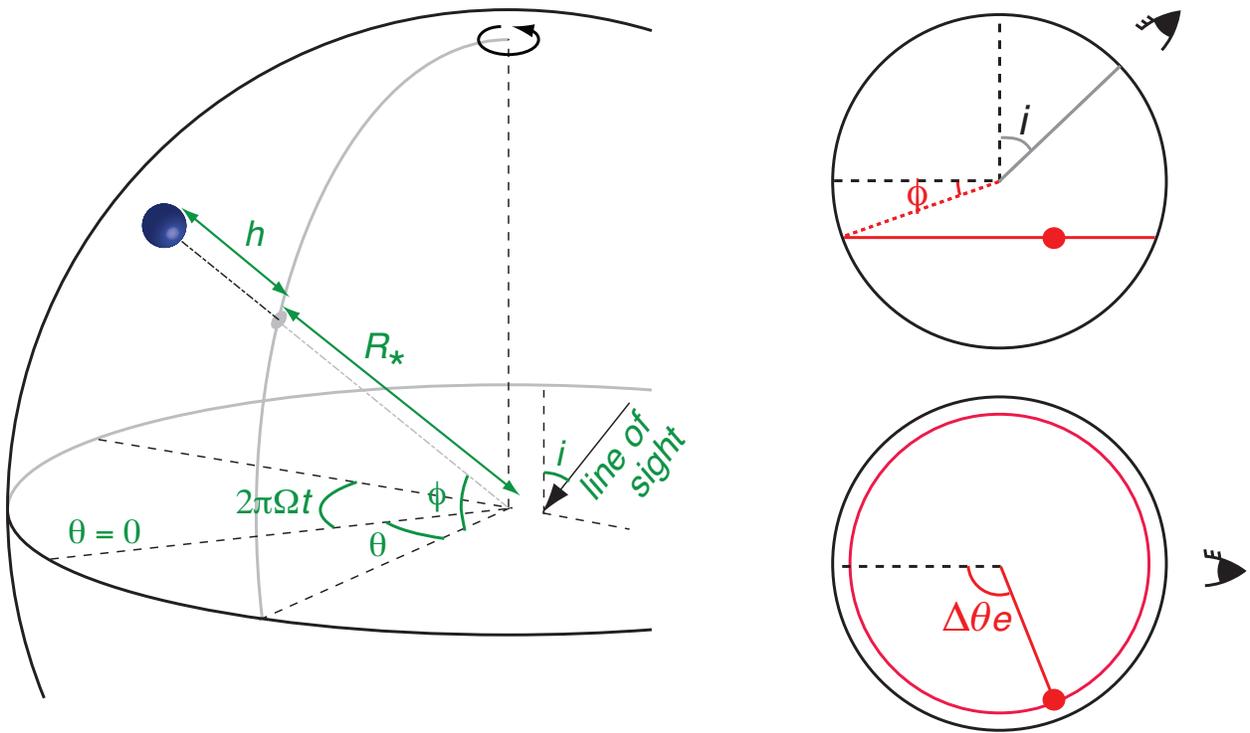}
\caption{Definitions of the coordinate system, $\theta$, $\phi$, $h$ and $i$ ({\it left}) and of $\Delta\theta_{e}$ when $h = 0$ ({\it right top}: edge-on view, {\it right bottom}: pole-on view).
\label{fig:geometry_point}
}
\end{figure}

\begin{figure}[t]
\plotone{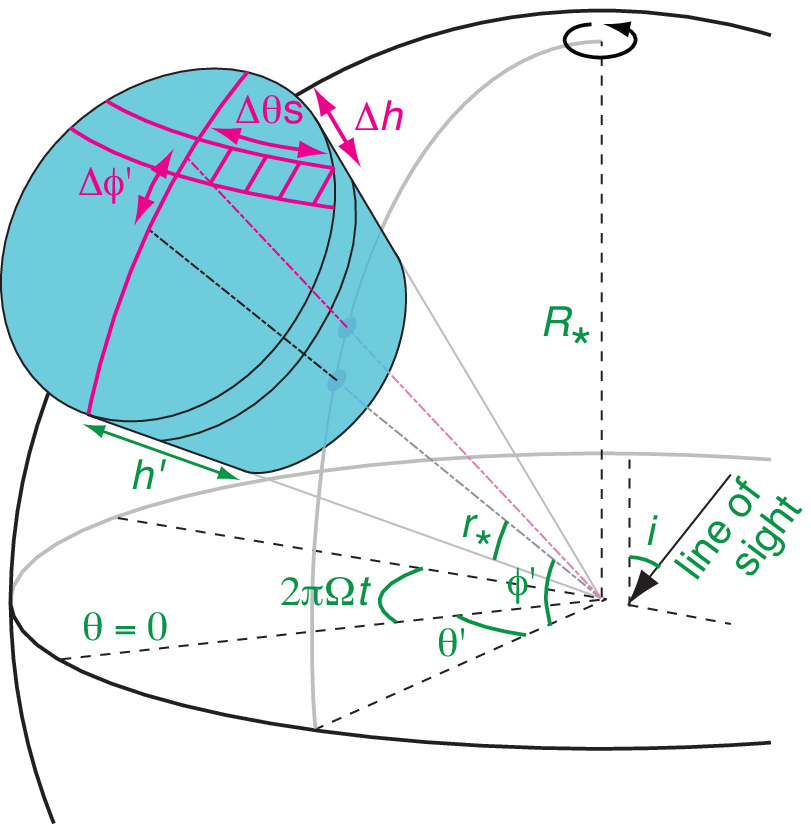}
\caption{Definition of the parameters $\theta'$, $\phi'$, $h'$, $\Delta\phi'$, $\Delta{h}$ and $\Delta\theta_{s}$
for the conically shaped plasma model.
\label{fig:geometry_cone}
}
\end{figure}

The plasma emitting volume is,
\begin{eqnarray}
V_{e} &=& \eta V\\
           &=& \eta\int_{0}^{h'} (R_{\ast}+h)^2\Omega dh\\
           &=& \frac{2\pi\eta(1-cos~r_{\ast})h'}{3}[3R_{\ast}^2+3R_{\ast}h'+h'^2]
\label{eqn:volume}
\end{eqnarray}
where $\eta$ and $\Omega$ are the plasma filling factor and the solid angle of the cone, respectively. 
To constrain the plasma density,
we combine the standard relation for plasma emission measure, $E.M. = n^2 V_{e}$ ($n$: plasma density),
with equation (\ref{eqn:volume}),
\begin{eqnarray}
\eta n^2 (\frac{R_{\ast}}{4R_{\odot}})^{3} &=& \frac{E.M.}{2 \pi (4R_{\odot})^{3} (1-cos~r_{\ast})h_{\ast}(1+h_{\ast}+h_{\ast}^2/3)}
\end{eqnarray}
where $h_{\ast} = h'/R_{\ast}$.
We calculate the right side of the equation for each combination of $r_{\ast}$ and $h_{\ast}$
and {\it E.M.} = $1.9\times10^{54}~cm^{-3}$ during the \SUZAKU\ observation in 2008 \citep{Hamaguchi2010}.
Fig.~\ref{fig:toymodel} {\it right} shows contours of values for this parameter.


\begin{deluxetable}{lccccc}
\tablecolumns{6}
\tablewidth{0pc}
\tabletypesize{\scriptsize}
\tablecaption{Analyzed Data Sets\label{tbl:obslogs}}
\tablehead{
\colhead{Abbreviation}&
\colhead{Observation ID}&
\colhead{Start Date}&
\colhead{Exposure}&
\colhead{Duration}&
\colhead{Net Count}\\
&&&\colhead{(ksec)}&\colhead{(ksec)}&\colhead{(counts)}
}
\startdata
\multicolumn{5}{l}{First Outburst:}\\
~~CXO$_{\rm 040307}$&5307&2004 Mar. 7&5.5&5.6&60\\
~~CXO$_{\rm 040322}\dagger$&5308&2004 Mar. 22&4.9&5.0&10\\
~~XMM$_{\rm 040404}$&0164560201&2004 Apr. 3&29.1&37.0&1321\\
~~XMM$_{\rm 050324}$&0301600101&2005 Mar. 24&79.2&89.7&1557\\
~~CXO$_{\rm 050411}$&5382&2005 Apr. 11&18.2&18.4&85\\
\multicolumn{5}{l}{Second Outburst:}\\
~~CXO$_{\rm 080918}$&9915&2008 Sept. 18&19.9&20.2&455\\
~~SUZ$_{\rm 081008}$&903005010&2008 Oct. 8&40.4&77.5&1275\\
~~CXO$_{\rm 081127}$&10763, 8585&2008 Nov. 27&20.0, 28.5&77.5&197,143\\
~~CXO$_{\rm 090123}$&9916&2009 Jan. 23&18.4&18.6&240\\
~~CXO$_{\rm 090421}$&9917&2009 Apr. 21&29.8&30.2&258\\
~~XMM$_{\rm 100228}\dagger$&0601960201&2010 Feb. 28&33.5&34.0&163
\enddata
\tablecomments{
Abbreviation: CXO --- \CHANDRA, XMM --- \XMM, SUZ --- \SUZAKU.
$\dagger$~The datasets are not used for the timing analysis.
Net count: Background subtracted photon counts between 1$-$8~keV.
Photon counts of all the available instruments are summed for \XMM\ and \SUZAKU.
}
\end{deluxetable}

\clearpage

\end{document}